\documentstyle[twoside,fleqn,espcrc2x,epsf]{article}

\input epsf

\newcommand{\AmS}{{\protect\the\textfont2
  A\kern-.1667em\lower.5ex\hbox{M}\kern-.125emS}}
\def\makeadmark#1{\hbox{$^{\rm #1}$}}

\def\ie{{\it {\frenchspacing i.{\thinspace}e. }}}
\def\eg{{\frenchspacing e.{\thinspace}g. }}

\def\simlt{\hbox{ \rlap{\raise 0.425ex\hbox{$<$}}\lower 0.65ex\hbox{$\sim$} }}
\def\simgt{\hbox{ \rlap{\raise 0.425ex\hbox{$>$}}\lower 0.65ex\hbox{$\sim$} }}

\def\that{{\hat t}}

\def\VEV#1{\left\langle #1\right\rangle}

\def\kpc{{\rm\, kpc}}

\def\ten#1{\times 10^{#1}} 

\def\msun {{ \rm \, M_\odot}} 
\def\mass {m} 

\def\Amax{A_{\rm max}} 
\def\umin{u_{\rm min}} 
\def\tmax{t_{\rm max}} 
\def\eff {{\cal E}} 
\def\ndf{N_{\rm dof}}
\def\Nobs{N_{\rm obs}}

\def\@versim#1#2{\lower0.2ex\vbox{\baselineskip\z@skip\lineskip\z@skip
  \lineskiplimit\z@\ialign{$\m@th#1\hfil##\hfil$\crcr#2\crcr\sim\crcr}}}
\def\apj{{\rm ApJ}}
\def\mnras{{\rm MNRAS}}
\def\pasp{{\rm PASP}}
\def\prl{{\rm Phys.~Rev.~Lett.}}
\def\aap{{\rm A\&A}}
\def\araa{{\rm ARA\&A}}
\def\nat{{\rm Nature}}
\def\iaucirc{\ref@jnl{IAU~Circ.}}
\def\yrone {A96} 
\title{The MACHO Project 2nd Year LMC Microlensing Results \\
and Dark Matter Implications}

\author{M.R.~Pratt\makeadmark{a,b},
\address{Astronomy, Box 351580, University of
Washington, Seattle, WA 98195-1580}
      \address{Center for Particle Astrophysics, University of California
        Berkeley, Berkeley, CA 94720}
C.~Alcock\makeadmark{b,c},
\address{Lawrence Livermore National Laboratory, Livermore, CA 94550}
R.A.~Allsman\makeadmark{d}
\address{Supercomputing Facility, Australian National University,\\
Canberra, ACT 0200, Australia},
D.~Alves\makeadmark{c,e}\address{Department of Physics, University of
California, Davis, CA 95616},
T.S.~Axelrod\makeadmark{f}
\address{Mount Stromlo and Siding Springs Observatories,
Australian National University,\\Weston, ACT 2611, Australia},
A.C.~Becker\makeadmark{a},
D.P.~Bennett\makeadmark{b,c,e},
K.H.~Cook\makeadmark{c,b},
K.C.~Freeman\makeadmark{f},
K.~Griest\makeadmark{g,b},
\address{Department of Physics, University of
        California San Diego, La Jolla, CA 92093-0350}
J.~Guern\makeadmark{g,b},
M.J.~Lehner\makeadmark{g,b},
S.L.~Marshall\makeadmark{b,c},
B.A.~Peterson\makeadmark{f},
P.J.~Quinn\makeadmark{h},
\address{European Southern Observatory, Karl-Schwarschild Str. 2, D-85748,
Garching, Germany}
A.W.~Rodgers\makeadmark{f},
C.W.~Stubbs\makeadmark{a,b},
W.~Sutherland\makeadmark{i}
\address{Department of Physics, University of Oxford,
Oxford OX1 3RH, U.K.}
and D.L~Welch\makeadmark{j}
\address{McMaster University, Hamilton Ontario Canada L8S 4M1}}
\begin{document}

\begin{abstract}
  The MACHO Project is searching for galactic dark matter in the form
  of massive compact halo objects (Machos).  Millions of stars in the
  Large Magellanic Cloud (LMC), Small Magellanic Cloud (SMC), and
  Galactic bulge are photometrically monitored in an attempt to detect rare
  gravitational microlensing events caused by otherwise invisible
  Machos.  Analysis of two years of photometry on 8.5 million stars in
  the LMC reveals 8 candidate microlensing events, far more than the
  $\sim1$ event expected from lensing by low-mass stars in known
  galactic populations.  From these eight events we estimate the
  optical depth towards the LMC from events with $2 < \that < 200$
  days to be $\tau_2^{200} \approx 2.9 ^{+1.4}_{-0.9} \ten{-7}$.  This
  exceeds the optical depth of $0.5\ten{-7}$ expected from known stars
  and is to be compared with an optical depth of $4.7\ten{-7}$
  predicted for a ``standard'' halo composed entirely of Machos.  The
  total mass in this lensing population is $\approx 2^{+1.2}_{-0.7}
  \ten{11} \msun$ (within 50 kpc from the Galactic center). Event
  timescales yield a most probable Macho mass of
  $0.5^{+0.3}_{-0.2}\msun$, although this value is quite model
  dependent.
\end{abstract}

\maketitle

\section{INTRODUCTION}
\label{sec-intro}
Galactic dark matter in the form of Machos can be detected by its
occasional gravitational microlensing of extragalactic
stars\cite{petrou81,pac86,griest91}.  This effect occurs when a Macho
of mass $\mass$ is in close alignment with a background star and acts
as a gravitational lens, magnifying and distorting the stellar image.
Microlensing refers to the situation in which image distortion is not
detectable ($\ll$~1~arcsec) and the only visible effect is an apparent
amplification, $\qquad A = (u^2 + 2)/u\sqrt{u^2 + 4}$.  Here $u =
b/r_E$ is the impact parameter in units of the Einstein radius $r_E =
\sqrt{4G\mass D_{OL}D_{LS}/c^2D_{OS}}$ where $D_{OL}, D_{LS}\: \&\:
D_{OS}$ are distances between observer, source and
lens\cite{refsdal64}.  This effect is transient with a timescale,
$\that \equiv 2r_E/v_\perp$, set by the mass and location of the lens
and its motion relative to the line of sight.  For a source in the
Magellanic Clouds and a lens at 10 kpc, $r_E \approx 8\sqrt{\mass
  /\msun}$ AU and $\that \approx 140\sqrt{\mass /\msun}$ days for
$v_\perp = 200$ km~s$^{-1}$.  With $\sim$daily observations over
periods of years, Machos in the range $10^{-4}-10^{2}\msun$ are
experimentally accessible.

Because the ``area'' of the lens is proportional to its mass, the
fraction of solid angle occupied by lenses, $\tau$,  depends only on
the mass density of Machos and not on their particular sizes.
Application of the virial theorem shows that
$\tau\!\sim\!(v_c/c)^2\!\sim\!5\ten{-7}$ for a Macho-dominated halo.
Here $v_c$ is the circular velocity about the center of the galaxy.
Thus detection of these very rare events requires many repeated
observations of millions of sources.  The optical depth towards the
Magellanic Clouds due to a Macho-dominated halo still comfortably
exceeds that from known populations of low-mass stars, so that
microlensing surveys directly probe the Macho content of the halo.

Several groups are now engaged in large scale photometric surveys to
search for dark matter in the form Machos using the high density of
stars in in the Magellanic Clouds, M31 and the Galactic bulge as a
backdrop for gravitational microlensing.  Four such surveys, DUO,
EROS, MACHO and OGLE, have detected in excess of 100 events that
appear to be microlensing
\cite{macho-nat,eros-nat,udalski93,udalski94b,bulge2,pratt95,alard96,lmc1,macho-prl}
of which $\sim$~10 are toward the LMC, the remainder being toward the
Galactic bulge.  Most of these events have characteristic timescales
$\that\sim20-60$ days.  Searches for short-timescale events with $1\ 
\mbox{hour} \simlt \that \simlt 10\ \mbox{days}$ have revealed no
candidates to date and set interesting limits on low-mass
Machos\cite{eros-ccd,spike,lehner96}.  Detailed reviews of
microlensing are given by \cite{pac-annrev} and \cite{roulet96}, a
basic outline of the MACHO project, and analysis procedures is given
in \cite{lmc1} (hereafter \yrone) and a more detailed analysis of this
sample is found in \cite{lmc2}.

\begin{figure}[thb]
\begin{minipage}[t]{7.5cm}
\epsfxsize=7.5cm 
\epsfysize=7.5cm 
\centerline{\epsffile{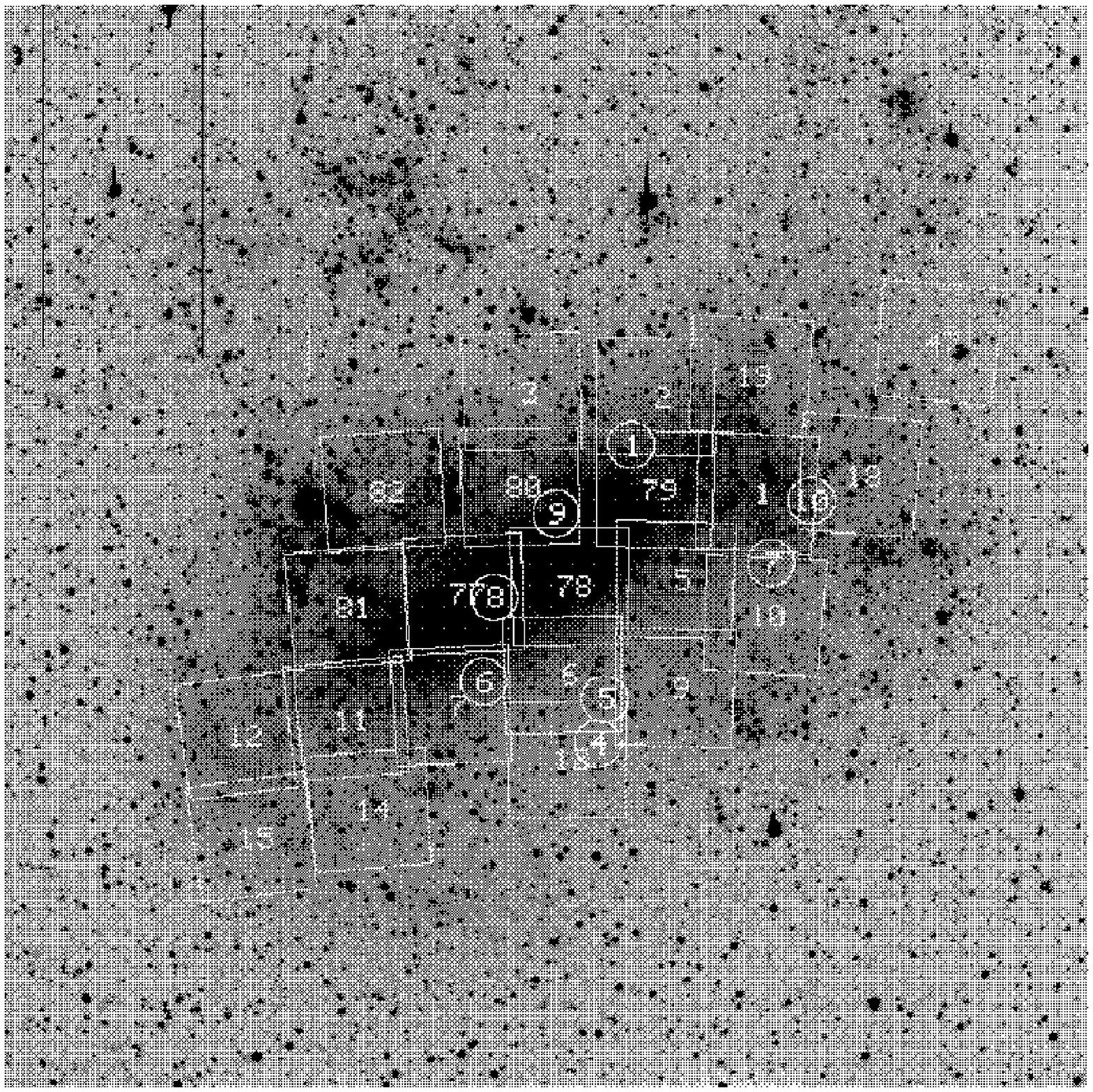}}
\label{fig-lmc}
{Figure \arabic{figure}. An R-band image of the LMC, $8.2$
  degrees on a side (G. Bothun, private communication), showing the
  locations of the 22 MACHO fields used here.  Event locations are
  are shown with circles.}
\end{minipage}
\end{figure} 

\section{OBSERVATIONS \& PHOTOMETRY}
\begin{table*}[hbt]
\setlength{\tabcolsep}{1.5pc}
\caption{Event selection criteria}
\label{tab-cuts}
\begin{tabular*}{\textwidth}{c|l|l}
\hline
& Description & Value \\
\hline
1 & Coverage & $\that < 300\,$days \quad \& $\quad \tmax \in$ data range \\
2 & Constant baseline & $\chi^2_{ml-out}/\ndf < 4$ with $\ndf > 40$ \\
3 & Amplification & $A_{\rm max} > 1+2\bar{\sigma}$ \\
4 & Significance & 6 measurements $> 1\sigma$ high \\
5 & Variable (Bumpers \& very red stars) 
    & $V > 17.5$ \quad \& \quad $V-R < 0.9$ \\ 
6 & SN 1987A light echos & 10'$\times$10' region excluded \\
7 & Peak fit & $\Delta\chi^2/(\chi^2_{\rm peak} / \ndf)>200$\\
8 & Crowding & $> 95\%$ of measurements have low crowding\\
9 & Crowding & $f_{CRD} < 70\log_{10}[\Delta\chi^2/(\chi^2_{ml} / \ndf)]-45$\\ 
10 & Significance & $\Delta\chi^2/(\chi^2_{ml} / \ndf) > 500$ \\
11 & Amplification & $A_{\rm max} > 1.75$ \\
\hline
\end{tabular*}
\end{table*}
 
\label{sec-obs}
The MACHO Project has had full-time use of the $1.27$-meter telescope
at Mount Stromlo Observatory, Australia, since mid 1992 in an extended
run until the end of the century.  Details of the telescope
system are given in \cite{macho-tel} and of the camera system in
\cite{stubbs93,marshall94}.  Briefly, corrective
optics and a dichroic beam splitter are used to simultaneously image a
$42\times42$ arcmin$^2$ field in two colors, using two mosaics of four
$2048^2$ pixel CCDs.
 
As of this writing, over 44000 exposures have been taken with the system,
over 3 TBytes of raw image data.  About $60\%$ are of the LMC, the
remainder are towards the Galactic bulge and SMC.  In this paper, we
consider only the first 2.1 years of data from 22 well-sampled LMC fields,
located in the central $\sim5^o\times3^o$ of the LMC and shown in
Figure~\ref{fig-lmc}.

The observations described here comprise 10827 images distributed over
the 22 fields. These include observations over a time
span of 840 days from mid 1992 to late 1994.  The mean number of
exposures per field is 492, with a range from 300 to 785.
The sampling varies between fields since the higher priority fields
were often observed twice per night with a $\sim 4$ hour spacing
for sensitivity to very short timescale events\cite{spike,lehner96}.

The photometric reduction procedure here is identical to that
described in \yrone; briefly, a good-quality image of each field is
chosen as a template, and used to generate a list of stellar positions
and magnitudes.  All other images are aligned with the template using
fiducial stars, and a PSF is determined from these. Then, the flux of
all other stars is fitted using the known positions and PSF. For each
measurement we also compute an error estimate and six quality flags;
these flags are the object type, the $\chi^2$ of the PSF fit, a
crowding parameter, a local sky estimate, and the fraction of the
star's flux rejected due to bad pixels and cosmic rays.  Suspect
measurements are removed based on these quality flags and the resulting
data are organized into stellar lightcurves.

\section{EVENT DETECTION}
\label{sec-events}
The MACHO Project is the largest survey of astronomical variability in
history and as such encompasses many previously unknown backgrounds to
microlensing.  The data used here comprise some 9 billion photometric
measurements.  Although we use objective selection criteria (cuts) to
discriminate genuine microlensing from stellar variability and
systematic photometry errors, these cuts are developed after extensive
examination of the MACHO database.  However, as long as the
experiment's event {\it detection efficiency} is calculated properly,
and the selection criteria are sufficiently stringent to accept only
real microlensing events, changes in the selection criteria will be
accounted for in the efficiency calculations, and the details will not
bias the final results.

Our experience with the Galactic bulge has been helpful in developing
the current cuts. Since our previous analysis of a smaller set of LMC
data (\yrone), we have analyzed a large set of data towards the
Galactic bulge \cite{bulge2,pratt95}, which has yielded over 80
microlensing events.  These observations provide a more realistic set
of microlensing events than the artificial events we previously used
to test our analysis procedures.  A number of these events fail the
original cuts because they are not well described by the ``normal''
microlensing light curve which assumes a single point lens, constant
velocities, and an unblended source star but nevertheless appear by
eye to be very high quality microlensing events. In contrast events 2
and 3 of \yrone\ appear far less striking.

This has motivated us to modify our selection criteria, shown in
Table~\ref{tab-cuts}, to have less stringent requirements on the
functional form of the lightcurve.  We compensate for this by
demanding higher signal to noise \ie higher amplification and greater
statistical significance.  

Event detection proceeds in several stages.  In the first stage, the
lightcurves of \S\ref{sec-obs} are convolved with 7, 15 and 45 day
matched filters as in \yrone\ and low level triggers are fit with the
``standard'' microlensing curve $A(u(t))$, where $u(t) = \sqrt{\umin^2
  + 4(t-\tmax)^2/\that^2}$.  We then compute for each light curve a
set of statistics and apply the cuts shown in Table~\ref{tab-cuts} to
distinguish microlensing from the background.  A measure of
significance is $\Delta\chi^2/(\chi^2_{ml}/\ndf)$, where $\Delta\chi^2
\equiv \chi^2_{const} - \chi^2_{ml}$ is the improvement in $\chi^2$
between a constant brightness fit and a microlensing fit.

\begin{figure}[thb]
\begin{minipage}[t]{7.5cm}
\epsfysize=7.5cm 
\centerline{\epsffile{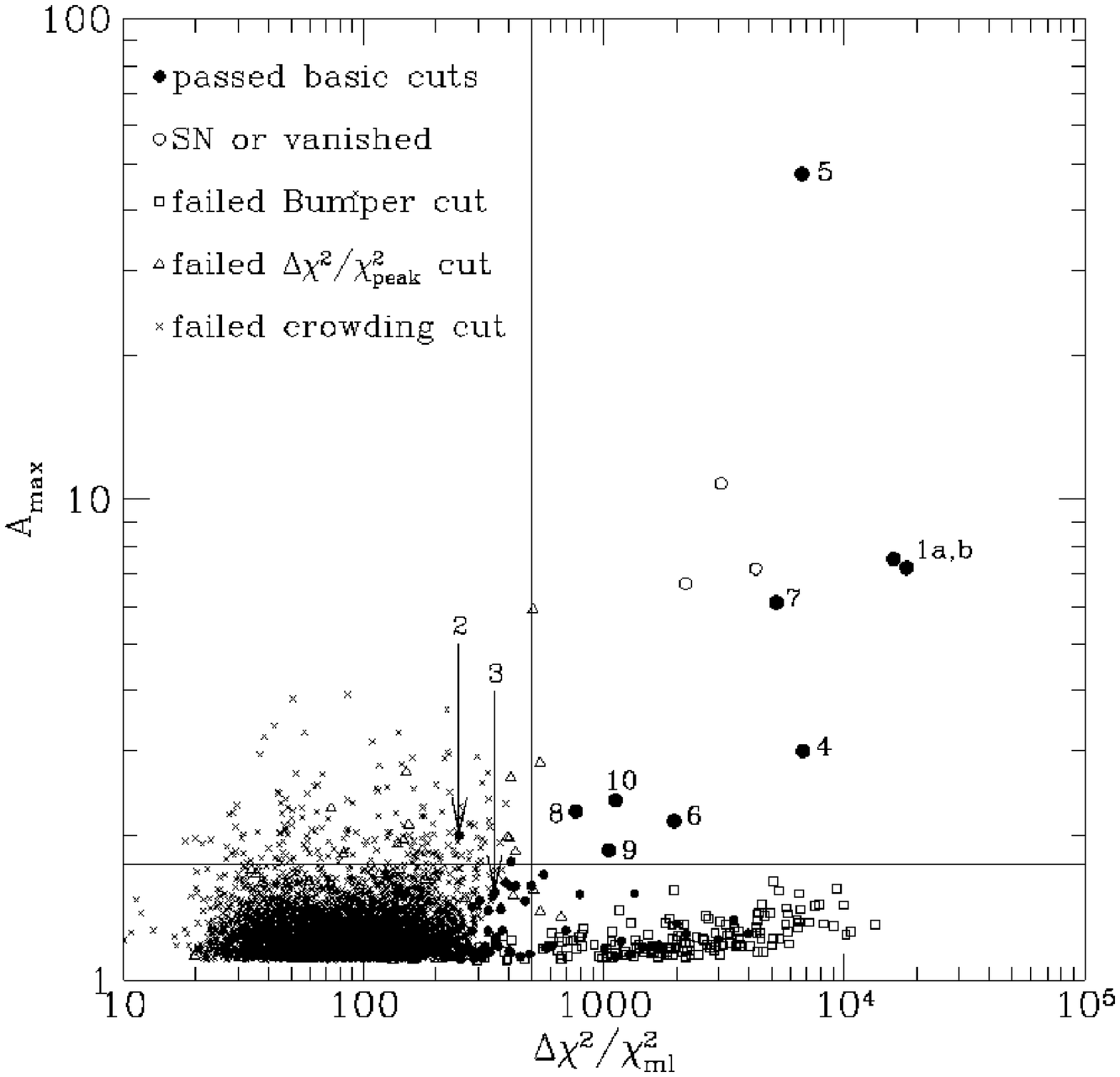}}
\refstepcounter{figure}
\label{fig-cuts}
{Figure \arabic{figure}. The final cuts for selection of
  microlensing candidates.   The y-axis is the fit $\Amax$.
  The solid lines show the final cuts 10 and 11; the circles in the
  upper right are the 12 events (10 stars) passing all cuts.}
\end{minipage}
\end{figure}

Two of the principal cuts are summarized in Figure~\ref{fig-cuts},
which illustrates cuts 10 and 11 for all events that pass cuts
1-4~\&~6.  Events that pass all cuts are indicated with circles,
while the other open symbols and crosses indicate events which failed
1 or more of cuts 5~\&~7-9.. Open squares indicate events which
fail cut 5 on magnitude, and open triangles indicate events which
fail cut 7 on peak fit. The crosses indicate events which fail
cuts 8 or 9.  The lightcurves passing all of these selection
criteria are further investigated, as outlined below.

\begin{figure}[thb]
\begin{minipage}[t]{7.5cm}
\epsfxsize=7.5cm 
\epsfysize=7.5cm 
\centerline{\epsffile{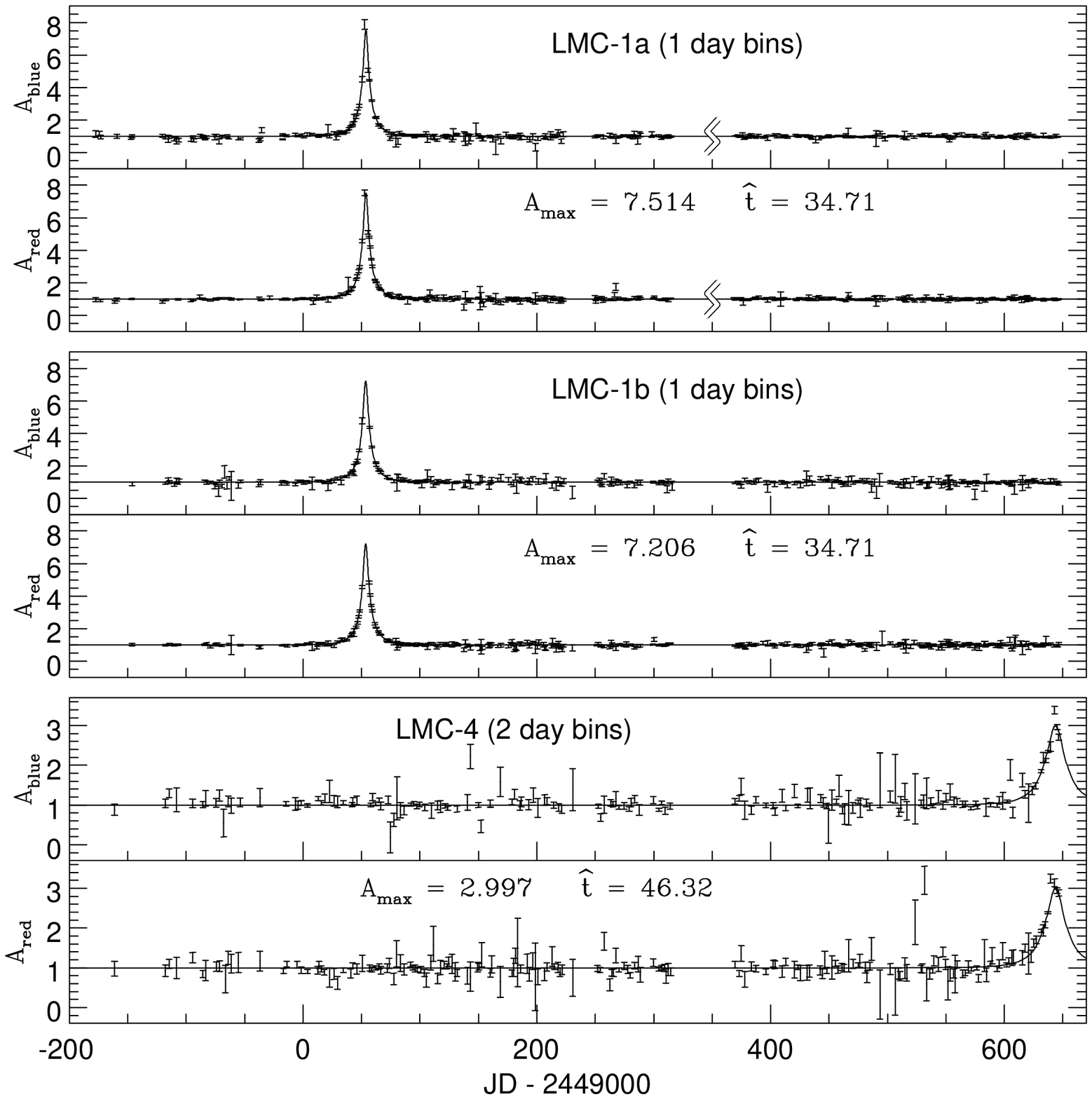}}
\refstepcounter{figure}
\label{events-figa}
{Figure \arabic{figure}. Candidate lightcurves 1 \& 4 }
\end{minipage}
\end{figure}

\section{MICROLENSING CANDIDATES}

\begin{figure}[thb]
\begin{minipage}[t]{7.5cm}
\epsfxsize=7.5cm 
\epsfysize=7.5cm 
\centerline{\epsffile{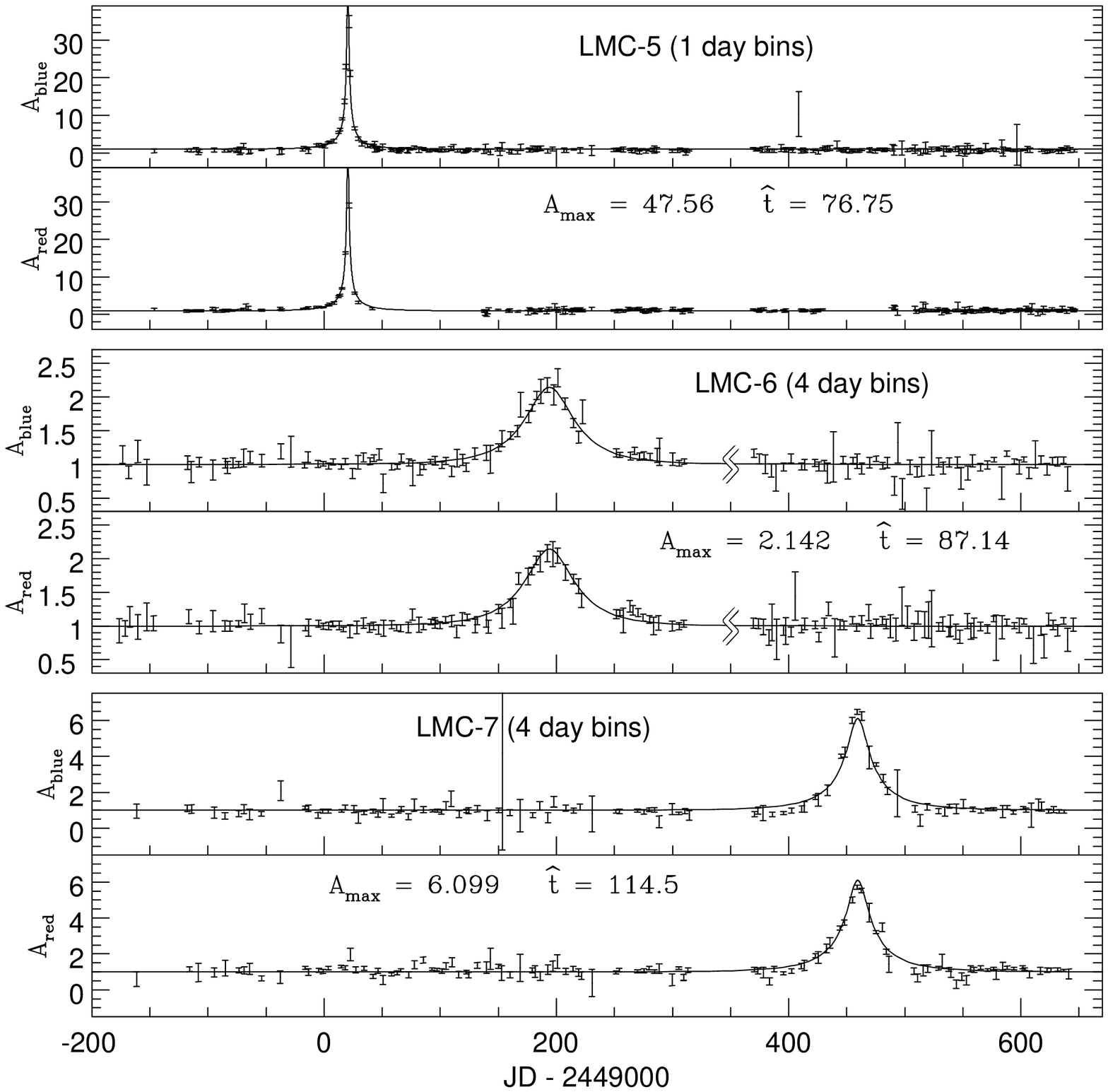}}
\refstepcounter{figure}
\label{events-figb}
{Figure \arabic{figure}. Candidate lightcurves 5-7 }
\end{minipage}
\end{figure}



Twelve lightcurves pass the cuts discussed above, indicated by the
large open and filled circles in the upper right hand region of
Figure~\ref{fig-cuts}.  Some lightcurves are shown in
Figures~\ref{events-figa}~\&~\ref{events-figb} and event fit parameters
are listed in Table~\ref{tab-events}.  Four of these lightcurves (1a,
1b, 12a and 12b) correspond to stars which occur in field overlap
regions; the two lightcurves for each star are based on independent
data and reductions and are detected independently.

Event~1 was our first discovery \cite{macho-nat}, and has remained
constant in following years.  The star's spectrum is that of a normal
clump giant at the radial velocity of the LMC \cite{dellavalle}.

The low signal-to-noise candidates 2 and 3 from \yrone\ {\bf do not
  pass} the final cuts used for this data set and are not included in
following calculations, but we have numbered the present set of 10
candidates $1, 4\ldots12$ to avoid possible ambiguity.  The EROS group
has informed us that the star involved in event~2 may have brightened
in 1990, and it has also shown indication of further brightening
episodes in our subsequent data; thus it is probably a variable star.

Event~4 was the first LMC microlensing candidate detected in progress
by our real-time Alert system\cite{pratt95} and announced in IAU
Circular 6095.  This event is confirmed by additional photometry and
spectra, obtained near peak amplification\cite{giraud}.

Event~9 passes our objective data cuts, but it appears to be lensing
by a binary system.  This event may resolve the extended size
of the source star, providing a measure of the relative proper motion
of the lens and source, which indicates the lens may reside in the LMC
\cite{bennett96}.

Event 10 passes all of our cuts, but our experience suggests that it
may be a variable star.  The asymmetry seen in the light curve, a
rapid rise with with a more gradual fall, is typical behavior one
expects from variable stars.

Events 11 and 12a/b are indicated with open circles in
Figure~\ref{fig-cuts} as they are only present in our reference
template images because of their amplification during that particular
observation.  They are no longer detectable in images that go somewhat
deeper.  Event~11 is superimposed on a background galaxy and is almost
certainly a supernova in that galaxy.  Event~12 may in fact be
microlensing.  Our efficiency analysis, described below, does not take
into account this ``amplification bias'' therefore it is necessary to
discard both of these events for the following calculations.

\begin{table*}[hbt]
\setlength{\tabcolsep}{1.5pc}
\setlength{\tabcolsep}{0.9pc}
\caption{Event parameters \& contribution to $\tau$}
\label{tab-events}
\begin{tabular*}{\textwidth}{rcccccc|cc} 
\hline
Event & V & V-R & $\tmax$ & $\that$ & $\Amax$ & $\chi^2/\ndf$
& $\that_{bl}$ & $\tau_1$\\
\hline
1a &   19.6 & 0.6  & 57.08(3)  & 34.7(3) & 7.2(1) & 1.420 & &\\
1b &   19.6 & 0.6  & 57.26(4)  & 34.3(3) & 7.5(3) & 1.134  
& \raisebox{1.5ex}[0pt]{38.8} &  \raisebox{1.5ex}[0pt]{$1.8 \ten{-8}$}\\
4 &  20.0 & 0.2  & 647.2(2) & 46(2) & 3.00(4)  & 1.416 
& 52 & $2.3 \ten{-8}$\\
5 &  20.7 & 0.4  & 24.0(3) & 82(2) & 58(5) & 1.680 & 88 
& $3.5 \ten{-8}$\\
6 &  19.6 & 0.3  & 197.5(7) & 87(4) & 2.14(4)  & 0.873 & 100 
& $4.1 \ten{-8}$\\
7 &  20.7 & 0.4  & 463.0(3) & 115(3) & 6.16(10) & 1.447 
& 131 & $6.0 \ten{-8}$\\
8 &  20.1 & 0.3  & 388.4(5) & 62(2) & 2.24(5)  & 2.218 & 70 
& $2.8 \ten{-8}$\\
9\footnote{Event~9 parameters are derived from a binary microlensing fit
\cite{bennett96}}
 &  19.3 & 0.3  &  603.04(2) & 143.4(2) & () & 1.755
& 143 & $6.6 \ten{-8}$\\
10 &  19.4 & 0.2  & 205.3(3)  & 42(1) & 2.36(5) & 1.982 & 47 
& $2.1 \ten{-8}$\\
11 &  21.5  & 0.4   & -8.6(3)  & 266(9) & 11.9(4) & 2.964 & &\\
12a &  21.2  & 0.3  & -10.0(3)  & 138(5) & 7.2(4) & 1.487 & &\\
12b &  21.2  & 0.3  & -11.4(8)  & 162(8) & 6.8(2) & 1.536 & &\\
\hline
\end{tabular*}
\end{table*}

These events, as shown in Figure~\ref{fig-lmc}, do not cluster spatially
or in any particular stellar population.  In addition, the
distribution of fit $\Amax$ is consistent with the expected uniform
distribution of $\umin$.  We classify events~1,4,5~\&~9 as `excellent'
microlensing candidates, events~6-8 as `good' and event~10 as
`marginal'. Events~11~\&~12 must be rejected as explained above, and
events~2~\&~3 from \yrone\ do not pass the current cuts.  We define
two sets of events for the following analysis.  The first set contains
all eight events passing the cuts without amplification bias.  The
second, more conservative, set of 6 events eliminates event~9 as a
possible LMC lens and event~10 as a possible variable star.

\section{DETECTION EFFICIENCY}
\label{sec-eff}

\begin{figure}[thb]
\begin{minipage}[t]{7.5cm}
\epsfysize=7.5cm 
\centerline{\epsffile{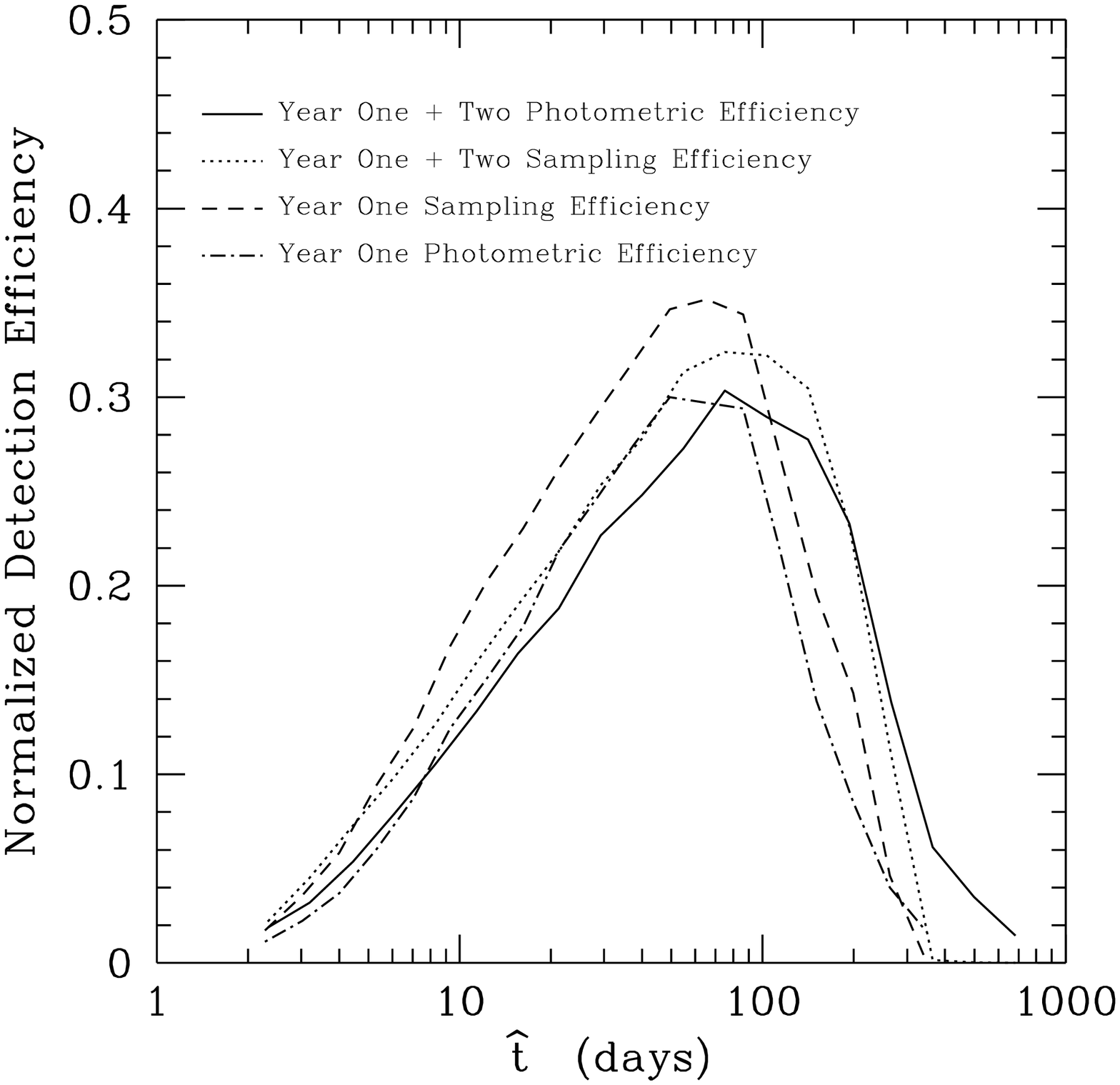}}
\refstepcounter{figure}
\label{fig-eff}
{Figure \arabic{figure}. Microlensing detection efficiency
  (normalized to $\umin < 1$) for the 2-year Macho data, as a function
  of event timescale $\that$.}
\end{minipage}
\end{figure}

The detection probability for individual events depends on many
factors, e.g. the 3 event parameters $\Amax, \that, \tmax$, and the
unlensed stellar magnitude, as well as our observing strategy and
weather conditions.  However, we can average over the known
distributions of $\Amax$ and $\tmax$ and stellar magnitude, using the
known time-sampling and weather conditions, to derive our efficiency,
shown in Figure~\ref{fig-eff}, as a function only of event timescale
$\eff(\that)$.

We have computed our detection efficiency using an essentially
identical method to that outlined in \yrone, simply generating
simulated microlensing events with $\that$ logarithmically distributed
in the range 0.3--1000 days and adding these simulated events
into the extended time-span of observations. The Monte-Carlo procedure
takes into account the actual spacing and error bars of the
observations, so any variations in sampling frequency, weather, seeing
etc.  between the first and second year data are automatically
accounted for.

Efficiency is defined as the fraction of input events with $\umin < 1$
which pass our cuts; since we use a cut of $\Amax > 1.75$ or $\umin <
0.661$, our efficiency is constrained to be less than 0.661.  This
efficiency is defined relative to an `exposure' of $E = 1.82 \ten{7}$
star-years, which arises from the 7.9 million well measured,
independent lightcurves in our sample multiplied by the 840-day span
of observations.

We calculate our efficiency with two levels of realism.  The first is
the `sampling' efficiency, in which we neglect stellar blending and
assume that all the additional flux in a microlensing event is
recovered by the photometry code. The second is the `photometric'
efficiency, where we add artificial stars to a representative set of
real images, re-run the photometry code to create look-up tables of
added vs.~recovered flux.  These look-up tables are then used to
generate artificial microlensing lightcurves in the same way as above.
This `photometric' efficiency is more realistic, and is typically
$\sim 20\%$ lower than the `sampling' efficiency for timescales less
than 200 days.

The photometric efficiency is based on all stars in our fields, even
those not uniquely identified because of S/N or crowding effects.
These are accounted for by integrating the detection efficiency per
star over a corrected luminosity function (LF) as in \yrone.  This
corrected LF is truncated about one magnitude beyond where our
measured LF becomes seriously incomplete.  However, the real LF
continues to rise as $10^{0.5m}$ for several magnitudes beyond this
cutoff so there should be an additional contribution to our exposure
from these fainter stars.  We have tried several different magnitude
cutoffs and it appears that our exposure ($E\eff$) is converging near
the cutoff used in this paper for events with $\that < 150\,$days.

In the photometric efficiency Monte Carlo, we find that in addition to
reducing our efficiency, extreme crowding tends to make events appear
lower amplitude and shorter.  This effect is due to ``blending'' or
the tendency of a group of several superimposed stars to be detected
as a single object.  We make statistical corrections for this effect
in column~8 of Table~\ref{tab-events}.  Our estimated systematic error
in the efficiency determination is $\sim10\%$, less than our
statistical error with eight events\cite{lmc2}.

The efficiency for timescales $\that \sim 10 - 60$ days is lower than
that from \yrone\ by $\sim 10$\% primarily because of the tighter
$\Amax$ and $\Delta\chi^2$ cuts.  For longer events, the two year data
is more sensitive because of its longer time span.

\section{OPTICAL DEPTH}
\label{sec-implic}
The simplest measurable quantity in a gravitational microlensing
experiment is the microlensing optical depth.  In the the limit
$\tau\ll1$, this is the instantaneous probability that a random star is
magnified by a lensing object by more than a factor of 1.34
($\umin<1$). This is related to the mass in microlensing objects along
the line of sight to the source stars by
\begin{equation} 
\label{eq-taudef}
 \tau = {4 \pi G \over c^2} \int \rho(l) {l (D_{OS}-l) \over D_{OS}} \, dl 
\end{equation} 
Thus, it depends only on the density profile of lenses, not on their
masses or velocities.  Experimentally, one can obtain an unbiased
estimate of the optical depth as
\begin{equation}
\label{eq-taumeas}
 \tau_{\rm meas} = {1 \over E} {\pi\over 4} 
                     \sum_i {\hat t_i \over \eff(\that_i)} \ .
\end{equation}
where $E$ is the total exposure (in star-years), $\that_i$ is the
Einstein ring diameter crossing time, and $\eff({\that_i})$ is the
detection efficiency.  Here, and below, we use the blend
corrected values, $\that_{bl}$, from column 8 of
Table~\ref{tab-events}.

Confidence levels on $\tau_{\rm meas}$ are determined from Monte Carlo
experiments, in which the number of events is a Poisson variable with
adjustable mean, and event timescales are drawn at random from the
observed set.  The mean is fixed so that 16\% of experiments yield an
optical depth larger/smaller than that measured.  The mean optical
depth for these experiments is then the ``$1\sigma$'' lower/upper
bound on $\tau_{\rm meas}$.

Using our full sample of 8 events, we find an optical depth for events
of duration $2$ days $< \that < 200$ days of $\tau_2^{200} = 2.9 {+1.4
  \atop -0.9} \ten{-7}$. If we subtract the predicted background
microlensing optical depth of $\tau_{\rm backgnd} = 0.5 \ten{-7}$
(from Table~\ref{tab-stars} below), we find that the observed excess
is about 50\% of the predicted microlensing optical depth for a
``standard'' all-Macho halo of \eg \cite{griest91}.  Alternatively, we
can estimate the optical depth due only to the halo by using the 6
event subsample defined in section \ref{sec-events}, for which
$\tau_{2}^{200} = 2.1 {+1.1 \atop -0.7} \ten{-7}$, about 45\% of the
optical depth predicted by a ``standard'' all-Macho halo.

Uncertainties can be much larger than those calculated above if our
measured distribution of timescales is a poor estimate of the real
timescale distribution, especially if very long events are possible.
It is worth noting that errors are considerably larger than Poisson
because they are dominated by the shot noise on the small number of long
events rather than the total number of events.  

\section{NON-HALO POPULATIONS}
\label{sec-stars}
\begin{table*}[hbt]
\setlength{\tabcolsep}{1.5pc}
\setlength{\tabcolsep}{0.9pc}
\caption{Microlensing by stars}
\label{tab-stars}
\begin{tabular*}{\textwidth}{rccccc} 
Population & $\tau (10^{-7}) $ & $ \VEV{\that}$ (days) & $ \VEV{l} (\kpc) $ 
& $ \Gamma (10^{-7} {\rm yr^{-1}}) $ & $ N_{\rm exp} $ \\
\hline
Thin disk   &  0.15  & 112  & 0.96  & 0.62  & 0.29  \\ 
Thick disk  & 0.036  & 105  & 3.0   & 0.16  & 0.075 \\
Spheroid    & 0.029  & 95   & 8.2   & 0.14  & 0.066 \\
LMC center  & 0.53   & 93   & 49.8  & 2.66  & (1.19) \\
LMC average & 0.32   & 93   & 49.8  & 1.60  & 0.71  \\
Standard halo & 4.7    & 89   & 14.4  & 24.3  & 11.2  \\
\hline
\end{tabular*}
\end{table*}

An in depth assessment of the microlensing background is 
necessary to determine if $\tau_{\rm meas}$ and $N_{\rm meas}$ are
significantly greater than that expected from non-halo populations.
As noted by \cite{wu94,sahu94,dgmr} and \yrone, low-mass stars in the
Galaxy and LMC will give rise to some microlensing events.  However,
these authors find that the optical depth from known stars is only
$\simlt 10\%$ of that from an all-Macho halo.

Shown in Table~\ref{tab-stars} are estimates of stellar lensing rates
of \yrone, using the same model parameters for the thin and thick
disks, the spheroid and the LMC disk, assuming a Scalo PDMF for all
populations, and simply updating the total exposure for our 2-year
dataset.  In our present sample, we expect a total background of
1.14 events from all know stellar populations.

A Poisson distribution with a mean of 1.14 gives probabilities to
observe $\Nobs \ge 3,4,5,6,7$ as $10.7\%, 2.9\%, 0.6\%, 0.11\%,
0.02\%$ respectively.  Thus we see that if only three of our events
are genuine microlensing, the evidence for an excess is modest,
whereas if more than four of our events are microlensing there is very
strong evidence for an excess over stellar lensing alone.

If one supposes that only the four ``best'' events result from lensing
by stars and the other candidates are not microlensing (e.g. they are
variable stars) then there is only marginal evidence for Machos in the
halo.  The flaw in this argument is that the distribution of
microlensing magnifications is given {\it a priori}, and the ``best''
microlensing candidates are preferentially the high magnification
ones.  We should expect to find a mixture of high and
low magnification candidates, and it is very unlikely that all events
would be high magnification given our detection efficiency.  By this
argument, the surplus of events is not critically dependent on the
lower quality candidates and it is difficult to explain
our results by stellar lensing alone.

\section{LIKELIHOOD ANALYSIS}
\label{sec-like}
\begin{figure}[thb]
\begin{minipage}[t]{7.5cm}
\epsfysize=7.5cm 
\centerline{\epsffile{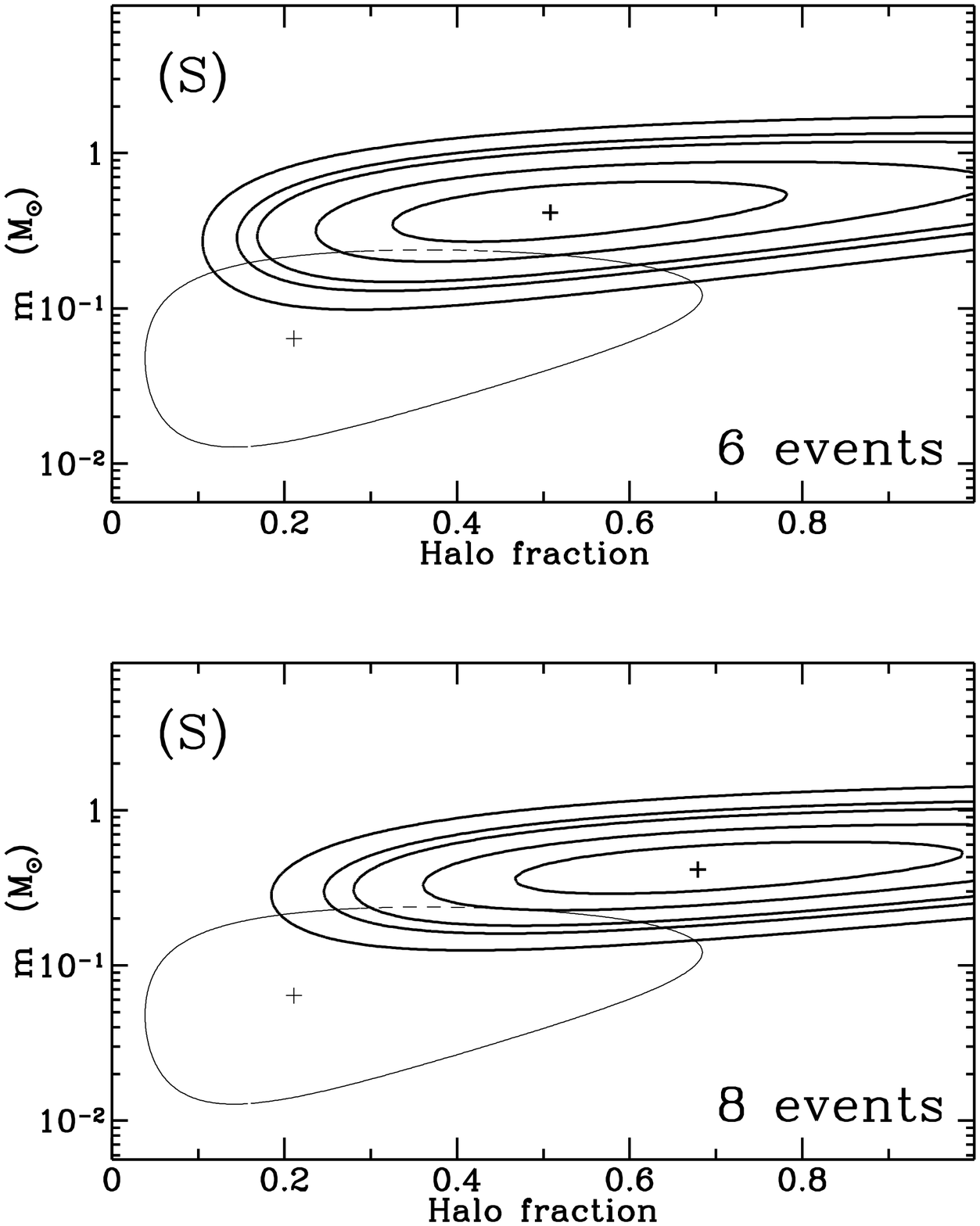}}
\refstepcounter{figure}
\label{fig-like}
{\vspace{0.2cm} Figure \arabic{figure}. Likelihood contours of
  Macho mass $m$ and halo Macho fraction $f$ for a delta-function mass
  distribution, for halo model S.  The most likely value is indicated
  with a $+$, and the contours enclose total probabilities of $34\%,
  68\%, 90\%, 95\%, 99\%$, using a Bayesian method as described in
  \S~\protect\ref{sec-like}.  The light line shows the $90\%$ contour
  from \yrone.}
\end{minipage}
\end{figure}

Since the timescale of a lensing event is proportional to $\sqrt{m}$,
we may use the observed event timescales to estimate the lens masses,
given a particular halo model to fix the phase space distribution of
lenses.  We apply a maximum-likelihood method as in \yrone\ using the
``standard'' halo where a fraction $f$ of the dark halo is made of
Machos with a unique mass $m$ (the remaining $1-f$ of the halo is
assumed to be unobserved dark matter).  The resulting likelihood
contours, assuming a delta-function mass function, are shown in
Figure~\ref{fig-like}; the probabilities are computed using a Bayesian
method with a prior uniform in $f$ and $\log\ m$.  The peak of the
likelihood contours gives the most probable mass and halo fraction for
this halo model.  For the 6 event sample we find $m_{2D} = 0.41
\msun$, and $f_{2D} = 0.51$.

We calculate one-dimensional likelihoods by integrating
over the other parameter.  For the six event sample this yields a most
likely Macho mass $m= 0.46^{+0.30}_{-0.17} \msun$, and most
likely halo fraction $f = 0.50^{+0.30}_{-0.20}$.  The errors
given are 68\% CL.  It is important to note the large extent of the
contours in Figure~\ref{fig-like}.  This is mostly due to the small
number of events.  The 95\% CL contour includes halo fractions from
17\% to 100\%, and Machos masses from 0.12 to 1.2$\msun$.

The most probable mass and halo fraction are both larger than our
results from year-1, which were $m = 0.06 \msun$ and $f = 0.2$, though
there is a reasonable degree of overlap of the contours.  The year-1
90\% CL contour is shown as the light line in Figure~\ref{fig-like}.
The most probable value of each analysis lies outside the 90\% CL
contour of the other analysis.  This is primarily due to the upward
shift in $\VEV{\that}$.  Events~2 \& 3 from \yrone, dropped in this
analysis, are both shorter than any events in the current set.  The
shift in $\VEV{\that}$ is also to be expected as our maximum efficiency
shifts toward longer events with the increasing baseline of the
experiment.

\section{DISCUSSION}
One natural explanation of the results presented here is that a
substantial fraction of the Galactic dark halo is in the form of
Machos.  The fact that the observed events have relatively
long timescales suggests that (for canonical halo models) the lenses
have masses above $\sim 0.1 \msun$, with a most probable mass $\sim
0.5 \msun$.  If so, they cannot be ordinary hydrogen-burning stars
since there are strong direct limits on such objects from counts of
faint red stars (e.g. \cite{bfgk,hhgc,fgb}); thus stellar remnants
such as white dwarfs appear to be an obvious possibility.

There are some theoretical difficulties with the white-dwarf
hypothesis (e.g. \cite{carr94,char-silk}). First, the initial mass
function must be fairly sharply peaked between $\sim 2 - 6 \msun$ to
avoid overproducing either low-mass stars (which survive to the
present) or high-mass stars (which as type-II SNe overproduce metals).
A second difficulty is that the high luminosity of the B and A
progenitors may exceed the observed faint galaxy counts.
 
However, white dwarfs are one of the few dark matter candidates which
are known to exist in large numbers.  Also, it has recently become
clear (e.g.  \cite{white93}) that the mass of hot gas in rich galaxy
clusters greatly exceeds that in stars, and furthermore this gas is
relatively metal-rich with an iron abundance $\sim 0.3$ solar.  This
might suggest that most of the baryons have been processed through
massive stars, which have since died leaving a population of remnants
and metal-rich gas. This scenario has been explored by \cite{mathews},
who suggest that it may be natural to have $40$-$100\%$ of the
Galactic dark matter in white dwarfs.

The observational limits on the local density of white dwarfs are a
strong function of their age \cite{liebert,chabrier}.  Very recently, a
limit from the Hubble Deep Field has been given by \cite{fgb}; they
find that white dwarfs with $M_I < 16$ contribute $< 100\%$ of the
halo density, and those with $M_I < 15$ contribute $< 33\%$.

Our exploration of halo models also shows that for `minimal' halos,
the timescales may still be consistent with substellar Machos just
below $0.1 \msun$. Also, models with a substantial degree of halo
rotation may lead to smaller mass estimates, since the rotation could
reduce the transverse component of the Macho velocities. Thus, brown
dwarfs cannot be ruled out as yet.

Finally, it is worth noting that microlensing is sensitive to any
compact objects, irrespective of composition, as long as they are
smaller than their Einstein radii.  Thus, more exotic objects such as
primordial black holes and strange stars are possible candidates that
are virtually undetectable by direct searches.

The most direct method for identifying the lens population is to
measure microlensing `parallax' by simultaneously observing events
from earth and a small telescope in Solar orbit \cite{gould-1sat}.
This measures the velocity of the lens projected to the Solar system,
which provides a definitive proof of microlensing, and discrimination
between disk, halo and LMC lenses on an event-by-event basis.

To summarize, our results indicate a microlensing optical depth of
$\approx 3 \ten{-7}$ or a Macho mass within 50 kpc of $\approx 2
\ten{11} \msun$. This provides evidence that Machos with masses in the
range $0.05 - 1 \msun$ contribute a substantial fraction of the
galactic dark halo.  Continued observations from this and other
projects should clarify this in the next few years.

\section{ACKNOWLEDGMENTS}
We are very grateful for the skilled support given our project by
S.~Chan and the technical staff at the Mt. Stromlo Observatory.  
Work performed at LLNL is supported by the DOE under contract
W-7405-ENG.  Work performed by the Center for Particle Astrophysics
personnel is supported by the NSF through AST 9120005.  The work at
MSSSO is supported by the Australian Department of Industry, Science,
and Technology., K.G. acknowledges support from DoE OJI, Alfred P.
Sloan, and Cotrell Scholar awards.  C.S. acknowledges the generous
support of the Packard, Sloan and Seaver Foundations.

\end{document}